\documentclass[pre,aps,amsmath,amssymb,reprint,floatfix,superscriptaddress]{revtex4-1}

\pdfoutput=1
\usepackage[T1]{fontenc}
\usepackage{bm}
\usepackage{graphicx}
\usepackage{xcolor}
\usepackage[colorlinks=true,citecolor=blue,urlcolor=blue]{hyperref}

\def\cal#1{\mathcal{#1}}
\def\av#1{\langle #1 \rangle}
\def\beq{\begin{equation}}
\def\eeq{\end{equation}}
\def\bea{\begin{eqnarray}}
\def\eea{\end{eqnarray}}

\let\cedilla\c
\def\eqq#1{Eq.~(\ref{#1})}
\def\eq#1{(\ref{#1})}
\def\f#1{Fig.~\ref{#1}}
\def\s#1{Section~\ref{#1}}
\def\c#1{~\cite{#1}}
\def\cc#1{~Ref.~\cite{#1}}

\def\x{{\bm x}}

\def\kB{k_{\rm B}}
\def\kt{\kB T}
\def\d{{\rm d}}
\def\xt{(\x,t)}
\def\nuu{{\bm \nu}}
\def\ff{{\bm f}}

\def\xii{{\bm \xi}}
\def\zetaa{{\bm \zeta}}

\begin{document}

\title{The thermodynamic freedom of a thermodynamic computer}

\author{Stephen Whitelam}
\email{swhitelam@lbl.gov}
\affiliation{Molecular Foundry, Lawrence Berkeley National Laboratory, 1 Cyclotron Road, Berkeley, CA 94720, USA}

\begin{abstract}

Thermodynamic computers are stochastic physical devices designed to perform calculations at the thermal energy scale. Their operation is constrained by the equations of stochastic thermodynamics, among which are a set of bounds, known as speed limits, that relate a thermodynamic computer's run time to its computational progress and the heat it dissipates. Using the Wasserstein speed limit we assess the thermodynamic efficiency of a simulation model of a thermodynamic computer trained to 
perform a standard machine-learning classification task. On this task the thermodynamic computer is as capable as a simple multilayer perceptron. We show that different inference protocols allow the computer to operate within 40\% of the thermodynamic limit of efficiency without loss of accuracy, or to perform inference increasingly rapidly at fixed accuracy and thermodynamic efficiency. These results indicate that a thermodynamic computer designed for a particular task retains considerable freedom in its thermodynamic operation.

\end{abstract}

\maketitle

\section{Introduction}

The emerging field of thermodynamic computing is concerned with doing calculations using physical devices that operate on energy scales close to that of thermal fluctuations\c{conte2019thermodynamic,hylton2020thermodynamic,wimsatt2021harnessing,aifer2024thermodynamic,melanson2025thermodynamic,aifer2025solving,whitelam2026training,whitelam2026nonlinear,whitelam2026generative,lockwood2026blueprint,rolandi2026energy,kaiser2021probabilistic,borders2019integer,landauer1961irreversibility,bennett1973logical,wolpert2019stochastic}. Such devices evolve according to stochastic equations of motion that in many cases can be closely approximated by the Langevin equation\c{sekimoto1998langevin,risken1996fokker}. For instance, the overdamped Langevin equation
\beq
\label{lang1}
\dot{\x}= \mu \ff(\x)  + \sqrt{2 \mu \kt} \,\xii(t)
\eeq
describes the dynamics of the thermodynamic computer of\cc{aifer2024thermodynamic}, which is built from small electrical circuits. The fluctuating degrees of freedom $\x=(x_1,\dots,x_N)$ correspond to the voltage states of the computer's $N$ units; the mobility parameter $\mu$ sets the basic timescale of the computer, about $10^{-6}$ s; $\ff  = -\nabla V(\x)$ is the force due to the programmable potential of the computer; $\kt$ is the thermal energy scale; and the Gaussian white noise $\xii(t)$ has correlations $\av{\xi_i(t)}=0$ and $\av{\xi_i(t)\xi_j(t')} = \delta_{ij}\delta(t-t')$. 

The same type of equation can describe other devices, such as those built from mechanical elements\c{dago2021information} or Josephson junctions\c{ray2023gigahertz,pratt2025controlled}. In these cases the degrees of freedom $\x$ represent displacements or phases, and the characteristic times are about $10^{-3}$ s and $10^{-9}$ s, respectively. The aim of thermodynamic computing is to design the potential $V(\x)$ of the computer so that the device in question performs a desired calculation, whether in\c{kaiser2021probabilistic,borders2019integer,aifer2024thermodynamic,melanson2025thermodynamic,rolandi2026energy, lockwood2026blueprint} or out\c{whitelam2026nonlinear,whitelam2026generative,whitelam2026training} of thermal equilibrium.

Devices that evolve stochastically are subject to the constraints of stochastic thermodynamics\c{seifert2012stochastic,ciliberto2017experiments,peliti2021stochastic,wolpert2024stochastic,rolandi2026energy}. Among these constraints are a set of bounds, known as speed limits, that relate the elapsed time of a stochastic process to its progress and thermodynamic cost\c{okuyama2018quantum,shiraishi2018speed,nakazato2021geometrical,van2023thermodynamic,sabbagh2024wasserstein}. For systems governed by \eqq{lang1}, the Wasserstein speed limit\c{aurell2012refined,dechant2019thermodynamic} states that
\beq
\label{sl}
\tau \;\geq\; \frac{{\cal W_2^2}(p(0),p({\rm f}))}{\mu \kB T \,\Sigma}.
\eeq
Here $\tau$ is the total duration of the stochastic process, which we may interpret, in the context of thermodynamic computing, as a program's runtime. The numerator sets a notion of computational progress. ${\cal W_2}$ is the $L^2$--Wasserstein distance between the initial distribution $p(0)=p(\x,0)$ and the final distribution $p({\rm f})=p(\x,\tau)$ of the computer's degrees of freedom, quantifying how far the system has moved in probability space: the larger is ${\cal W_2}$, the more computational progress has been made. For \eqq{lang1}, the time evolution of the probability density $p(\x,t)$ is governed by the Fokker--Planck equation
\beq
\label{fp1}
\partial_t p = -\nabla\cdot\!\big(\nuu p\big), \quad \nuu = \mu(\ff - \kt \nabla \ln p),
\eeq
with initial condition $p(\x,0)$. Here $\nuu \xt$ is the velocity current.

The denominator of \eqq{sl} expresses the thermodynamic cost of the computation: it contains the mobility parameter, the thermal energy scale, and the total entropy produced by the computer,
\beq
\label{sig}
\Sigma = \int_0^{\tau} \!\d t\, \sigma(t), \quad
\sigma(t) = \frac{1}{\mu \kt} \int \d\x \,\| \nuu\|^2 p \geq 0.
\eeq
Here $\sigma(t)$ is the instantaneous rate of entropy production, and $\|\nuu\|^2 = \sum_{i=1}^N \nu_i(\x,t)^2$ is the squared Euclidean norm of the velocity current.

Here we use the speed limit \eq{sl} to assess the efficiency of a nonlinear thermodynamic computer trained to classify MNIST\c{lecun1998gradient} images. In \s{sec:model} we describe the computer, which is trained to operate at a predetermined observation time\c{whitelam2026training,whitelam2026nonlinear}. In \s{sec:dynamics} we show that a single observation of a single nonequilibrium trajectory per image gives an accuracy comparable to that of a multilayer perceptron; no repetition or averaging is required. In \s{sec:thermo} we explore the thermodynamic efficiency of the computer under different operating protocols. Gradual loading allows an increase of thermodynamic efficiency without sacrificing accuracy, to within about 40\% of the thermodynamic bound. The cost of increased efficiency is a reduction of the rate of computation. Alternatively, rescaling the potential of the computer and adding external noise\c{whitelam2025increasing,basak2026adding} accelerates the computation at fixed thermodynamic efficiency and entropy production, at the cost of increased heat emission. These results show that a computer performing a given task can do so in a range of thermodynamic regimes. 

\begin{figure*}[t]
\centering
\includegraphics[width=\linewidth]{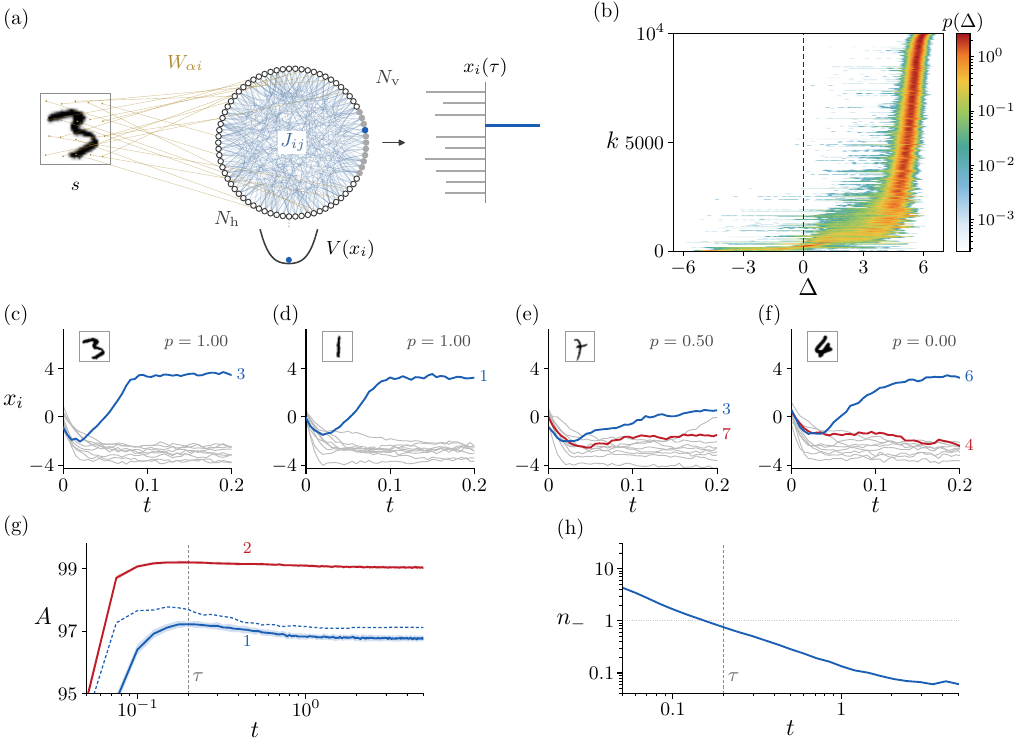}
\caption{Dynamics of inference of our model thermodynamic computer. (a) A 784-pixel MNIST image $s$ is connected to the computer via the couplings $W_{\alpha i}$. The $N=74$ units of the computer $x_i$ (divided into $N_{\rm h}=64$ hidden and $N_{\rm v}=10$ visible units) interact via the potential \eq{pot}, and evolve via the Langevin dynamics \eq{lang1}. The computer's predicted class $c$ corresponds to the output unit with the largest value $x_c(\tau)$ at time $t=\tau=1/5$ (times are measured in units of $\mu^{-1}$). (b) Probability distribution of the classification gap $\Delta$, \eqq{delta}, for all $10^4$ test-set digits, calculated using $10^3$ trajectories per digit. Classification is correct when $\Delta >0$. The index $k$ runs over the test images in order of increasing mean gap. (c--f) The computer's ten outputs, for a single trajectory for each of 4 images, as a function of time. Here $p$ is the probability (taken over $10^3$ trajectories) that the image is correctly classified. (g) Mean MNIST test-set classification accuracy (top-2 in red and top-1 in blue) as a function of time $t$; the device is trained to operate at $t=\tau=1/5$ (vertical dashed line). The faint colors indicate the standard deviation taken over $20$ realizations of the test set; the dashed blue line is the accuracy associated with the noise-free limit $\kt=0$. (h) The mean number $n_-$ of locally unstable directions of the computer's energy landscape, calculated from the eigenvalues of the Hessian, as a function of time.}
\label{fig1}
\end{figure*}

\section{A model thermodynamic computer}
\label{sec:model}

Our simulation model of a thermodynamic computer, sketched in \f{fig1}(a), consists of $N=74$ classical, real-valued degrees of freedom or {\em units} $\x=(x_1,\dots,x_N)$. Of these, 64 are hidden units and 10 are outputs. Units are subject to the potential energy
\bea
\label{pot}
V(\x)&=&\sum_{i=1}^N(J_2x_i^2+J_4x_i^4)-\sum_{i=1}^N b_ix_i\nonumber\\
&&+\sum_{i<j}J_{ij}x_ix_j+\sum_{i=1}^N\sum_{\alpha=1}^{N_{\rm in}}W_{\alpha i}s_\alpha x_i.
\eea
The first term in \eq{pot} is the on-site potential of each unit. The quartic term in $J_4$ gives each unit a nonlinear response to input forces, allowing the computer as a whole to perform calculations that are not linearly separable\c{whitelam2026nonlinear}. We set $J_2=J_4=\kt$. 

The coefficients $b_i$, $J_{ij}$, and $W_{\alpha i}$ are the adjustable parameters of the computer, and encode the desired program. In this paper we consider a computer trained, using methods described in Refs.\c{whitelam2026training,whitelam2026nonlinear}, to classify MNIST images\c{lecun1998gradient}. The $b_i$ are on-unit biases. The $J_{ij}$ are symmetric inter-unit couplings connecting all $N(N-1)/2$ pairs of units ($J_{ij} = J_{ji}$ and $J_{ii}=0$). The input couplings $W_{\alpha i}$ connect the $N_{\rm in}=28^2$ pixels of image $s$ to the computer's $N=74$ units; $s_\alpha$ denotes pixel $\alpha$ of image $s$. 

Inference proceeds as follows. Initially, all adjustable parameters are set to zero, and each unit is allowed to equilibrate independently in its on-site potential. At $t=0$ the adjustable parameters are set to the values identified by training, and $s \in \{1,2,\dots,10000\}$ is chosen to be one of the $10^4$ MNIST test-set images. The computer then evolves according to \eqq{lang1}, until $t=\tau=1/5$ (we measure times in units of $\mu^{-1}$). At this time we observe the values of the 10 output units of the computer. The output unit with the largest value $x_c(\tau)$ corresponds to the computer's prediction for the class $c \in \{0,1,\dots,9\}$ of the image.

\section{Dynamics of inference}
\label{sec:dynamics}

In this section we assess the dynamical operation of our model thermodynamic computer. The computer is designed to do classification using single dynamical trajectories of duration $\tau$. However, it is a stochastic device, and so its reliability has to be assessed using repeated realizations of the full test set. 

Using one trajectory for each image, the computer classifies the standard MNIST test set at an accuracy of $97.25 \pm 0.09\%$ (the mean and standard deviation determined using $10^3$ realizations). This accuracy is similar to that of a simple multilayer perceptron\c{lecun1998gradient}. The quartic term proportional to $J_4$ provides essential computational power: in its absence, the thermodynamic computer is a linear model that is considerably less capable. Training a thermodynamic computer with $J_4=0$ results in an accuracy of $90.5\%$, similar to that of a single linear multiclass rule.

Thermal noise influences {\em which} images the computer classifies correctly, as shown below, but the overall test-set accuracy does not vary significantly (standard deviation 0.09\%) upon repeated operation. Thus the training process has learned to accommodate thermal noise to the extent that the operation of a stochastic device is, at the level of overall accuracy, essentially predictable. (In the deterministic limit, when the computer is run at $T=0$, its classification accuracy is $97.70\%$, only $0.45$ percentage points above its value at $\kt=1$. However, a physical realization of the model described here would operate in the presence of thermal noise.)

To explore the computer's operation in more detail, we show in \f{fig1}(b) the probability distribution of the difference 
\beq
\label{delta}
\Delta \equiv x_{c^\star}(\tau)-\max_{c\ne c^\star}x_c(\tau)
\eeq
between the output associated with a given image's class $c^\star$ and the largest of the other outputs. A value of $\Delta>0$ indicates that the image has been correctly classified. Over $10^3$ trials per image, most images (about $86.3\%$) are always correctly classified. A small number, about $0.45\%$, are always misclassified. The remaining $13.2\%$ are sometimes classified correctly and sometimes misclassified, although they are heavily weighted toward correct classification: their mean success probability is $82.6\%$.

In \f{fig1} panels (c--f) we show dynamical trajectories of the computer's outputs when fed 4 different images. Panels (c) and (d) show two digits, a 3 and a 1, that are always correctly classified ($p$ is the probability of correct classification over $10^3$ trials per image). Panel (e) shows a 7 that is misclassified in half of all runs. Panel (f) shows a 4 that is almost always misclassified.

The computer is designed to provide output at a predetermined observation time $\tau=1/5$. At this observation time the computer happens to be out of equilibrium, as shown in \f{fig1}(g) and (h). Panel (g) shows the test-set accuracy (top-2 in red and top-1 in blue) if the computer's output were read at time $t$; the vertical dashed line denotes the time $\tau=1/5$ at which the computer is designed to be read. Accuracy peaks at $t=\tau$, and diminishes at shorter and longer times. The computer is still functional at longer times, when it reaches steady state (which may correspond to thermodynamic equilibrium or to a long-lived kinetically trapped state), but it is most accurate at the time at which it is trained to operate.

Panel (h) confirms that the computer, observed at its intended readout time of $\tau=1/5$, is out of equilibrium. The figure shows the mean number $n_-$ of negative eigenvalues of the Hessian 
\beq
H_{ij}(\x)=\partial_i \partial_j V(\x) = J_{ij}+\left(2J_2+12J_4x_i^2\right) \delta_{ij},
\eeq
 where $\delta_{ij}$ is the Kronecker delta. At $t=0$ the computer has about 34 locally unstable directions. This number decreases continuously as the computer runs, including at the designated readout time $t=\tau$. The quantity $n_-$ is therefore not stationary at readout time, and so the computer is out of equilibrium.

\section{Stochastic thermodynamics of inference}
\label{sec:thermo}

We quantify the thermodynamic efficiency of inference by writing the speed limit \eqq{sl} in the form
\beq
\label{eta}
\eta(t) \equiv  \frac{\overline{{\cal W_2^2}(p(\x,0),p_s(\x,t))}}{\mu t \, \kt \, \overline{\Sigma_s(t)}} \leq 1.
\eeq
We can interpret $\eta$ in \eqq{eta} as the ratio of the minimum runtime allowed by thermodynamics to the actual runtime of the computer, or as the ratio of the minimum entropy production allowed by thermodynamics to the actual entropy production of the computer. The most efficient computer possible has $\eta=1$. 

To calculate the efficiency $\eta$ we average thermodynamic quantities over noise (and so consider the dynamical ensemble generated by each image $s$), and subsequently average over data $s$. Thermodynamic quantities therefore carry an explicit data label $s$. The overbar $\overline{(\cdot)}$ denotes an average over data. The ratio of averages in \eq{eta} measures the efficiency of the computer's workload, and is the data-averaged transport divided by the data-averaged thermodynamic cost.

We consider two operating protocols: the first is a generalization of the procedure of \s{sec:dynamics} to allow the trained couplings of the computer to be turned on at finite rate. The second involves the controlled addition of noise to a computer whose couplings are uniformly rescaled versions of their trained values, following the prescription of\cc{whitelam2025increasing}. The first strategy allows increased efficiency $\eta$ at the cost of decreased rate of computation, while the second allows increased rate of computation at fixed efficiency, at the cost of increased heat emission.

\subsection{Under a loading protocol}

We modify the inference procedure of \s{sec:dynamics} in order to allow loading of the computer at finite rate. As before, the units of the computer are first allowed to equilibrate in their on-site potential 
\beq
\label{visolated}
V_0(\x)=\sum_{i=1}^N(J_2x_i^2+J_4x_i^4).
\eeq
The computer is then loaded over a time $t_\ell$ according to the protocol
\beq
\label{loadpot}
V_s(\x,t)=V_0(\x)+\Gamma_\ell(t)\Delta V_s(\x),
\eeq
where
\beq
\Gamma_\ell(t)=
\begin{cases}
1, & t_\ell=0,\\
\min(t/t_\ell,1), & t_\ell>0,
\end{cases}
\eeq
and
\beq
\label{vload}
\Delta V_s(\x)={}-\sum_{i=1}^Nb_ix_i+\sum_{i<j}J_{ij}x_ix_j 
+\sum_{i=1}^N\sum_{\alpha=1}^{N_{\rm in}}W_{\alpha i}s_\alpha x_i
\eeq
contains the image data and the inter-unit interactions. The instantaneous quench of \s{sec:dynamics} corresponds to $t_\ell=0$. During loading, the computer obeys the equation of motion
\beq
\label{langload}
\dot{\x}=-\mu\nabla V_s(\x,t)+\sqrt{2\mu\kt}\,\xii(t).
\eeq

To compute the required pieces of \eqq{eta} we proceed as follows.

We first calculate the denominator of \eq{eta}. Given input data $s$, the computer's entropy production is\c{seifert2005entropy}
\beq
\label{loadsigmamain}
\kt \Sigma_s(t)=\av{Q_s(t)}
+T \left( S[p_s(\x,t)]-S[p(\x,0)] \right).
\eeq
Here $Q_s(t)$ is the heat emitted to the bath, the average $\av{\cdot}$ is taken over independent dynamical trajectories, $p_s(\x,t)$ is the distribution of the computer's degrees of freedom $\x$ at time $t$, and $S$ is the Gibbs-Shannon entropy
\beq
\label{entropydef}
S[p]\equiv-\kB\int\d\x\,p(\x)\ln p(\x).
\eeq

Using the sign convention of\cc{sekimoto1998langevin}, the heat emitted to the bath is
\beq
\label{heat}
Q_s(t)=W_s(t)
-\left[V_s(\x(t),t)-V_0(\x(0))\right],
\eeq
where
\beq
\label{work}
W_s(t)= \int_0^t\!\d u\,\dot\Gamma_\ell(u)\Delta V_s(\x(u))
\eeq
is the work done by the loading protocol along a trajectory. For a quench, \eq{work} reduces to $W_s(t)=\Delta V_s(\x(0))$.

Prior to loading, the computer is in equilibrium with respect to the noninteracting energy function $V_0(\x)$, independent of input data, and so
\beq
\label{loadinitialentropy}
S[p(\x,0)]
=-N\kB\int\d x\,p_0(x)\ln p_0(x).
\eeq
Here $p_0(x)=z_0^{-1}\exp[-(J_2x^2+J_4x^4)/\kt]$ is the normalized equilibrium density of an isolated unit, with
\beq
z_0=\int\d x\,\exp[-(J_2x^2+J_4x^4)/\kt].
\eeq
For $J_2=J_4=\kt$ and $N=74$, we get $S[p(\x,0)] \approx 50.4 \,\kB$.
\begin{figure*}[t]
\centering
\includegraphics[width=\linewidth]{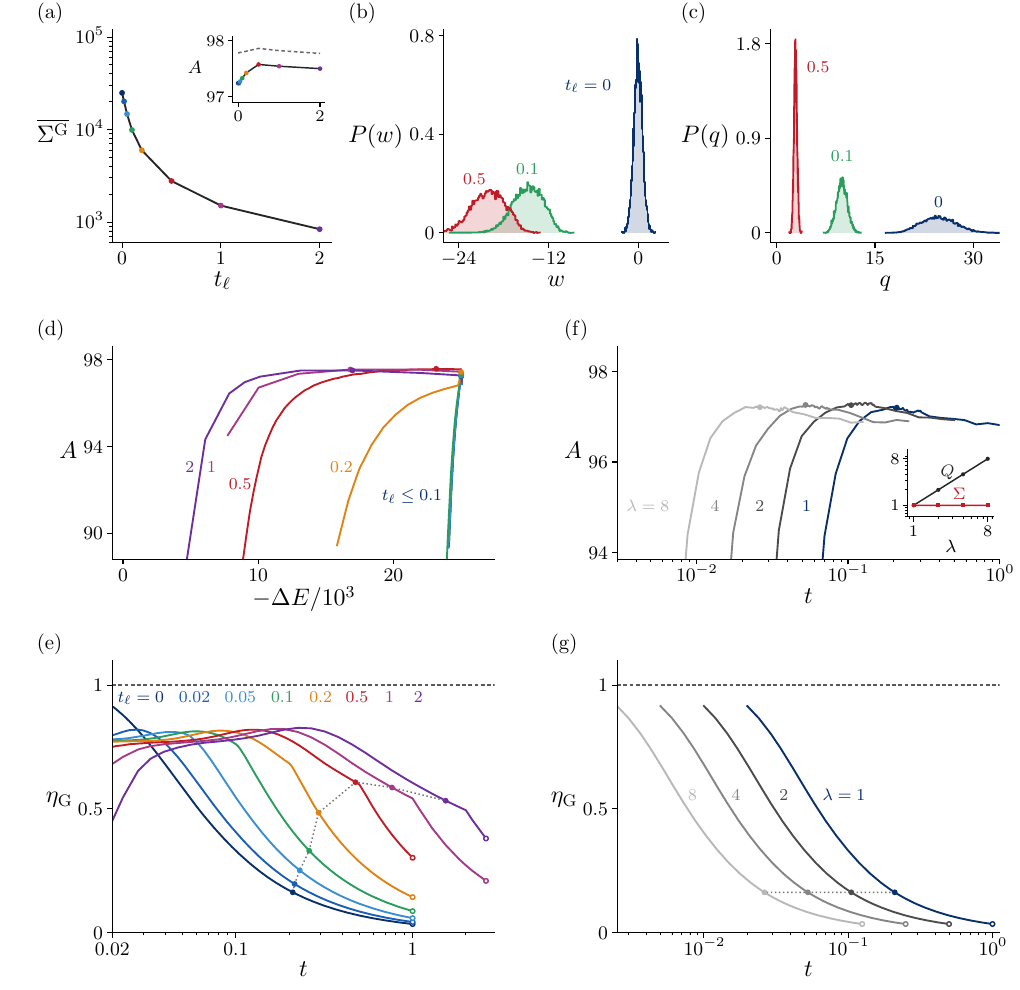}
\caption{Thermodynamics of inference. (a) Mean entropy-production estimate (at the most accurate readout time) as a function of loading time. Inset: the corresponding MNIST test-set classification accuracy, with the $T=0$ result shown dashed. Colors denote the loading times $t_\ell=0,0.02,0.05,0.1,0.2,0.5,1,$ and $2$ (from left to right). (b,c) Distributions of work and heat, with $w\equiv W/(10^3 \kt)$ and $q\equiv Q/(10^3 \kt)$; colors denote $t_\ell=0,0.1,$ and $0.5$. (d) Test-set accuracy as a function of the mean energy (in $\kt$) lost by the computer; colors denote the loading times listed in panel (a). (e) The efficiency bound $\eta_{\rm G}(t)$ under loading. Colors denote the loading times listed in panel (a), while dots indicate the readout times at which accuracy is greatest. The dotted line is a guide to the eye. (f) Accuracy under the clock-acceleration procedure. Colors denote $\lambda=1,2,4,$ and $8$, and dots indicate the observation times $t^*/\lambda$. Inset: heat and entropy production at the indicated readout time, relative to their values at $\lambda=1$. (g) The efficiency bound under the clock-acceleration procedure. Colors denote the values of $\lambda$ listed in panel (f), while dots indicate the observation times $t^*/\lambda$. The dotted line is a guide to the eye.}
\label{fig2}
\end{figure*}

We estimate the entropy of the finite-time distribution $p_s(\x,t)$ by calculating its mean and covariance,
\bea
\label{loadmoments}
\bm m_s(t)&=&\int\d\x\,\x\,p_s(\x,t),\nonumber\\
C_s(t)&=&\int\d\x\,
[\x-\bm m_s(t)][\x-\bm m_s(t)]^{\rm T}p_s(\x,t).
\eea
We then calculate the Gaussian entropy
\beq
\label{loadgaussianentropy}
S_{\rm G}[p_s(\x,t)]
=\frac \kB 2\ln[(2\pi{\rm e})^N\det C_s(t)].
\eeq
A Gaussian has the largest entropy at fixed covariance, and so provides an upper bound on the true entropy:
\beq
S[p_s(\x,t)]\leq S_{\rm G}[p_s(\x,t)].
\eeq
Replacing the true finite-time entropy in \eq{loadsigmamain} by its Gaussian bound \eq{loadgaussianentropy} gives us a bound on the entropy production:
\beq
\label{loadsigmagaussian}
\Sigma_s^{\rm G}(t)\equiv
\frac{\av{Q_s(t)}}{\kt}
+\frac{S_{\rm G}[p_s(\x,t)]-S[p(\x,0)]}{\kB}
\geq\Sigma_s(t).
\eeq

We next calculate the numerator of \eq{eta}, the $L^2$--Wasserstein distance between $p(\x,0)$ and $p_s(\x,t)$. This distance is the minimum mean-squared displacement required to transform one distribution into the other\c{dechant2019thermodynamic,nakazato2021geometrical,sabbagh2024wasserstein}:
\beq
{\cal W}_2^2(p,p')
\equiv \inf_{\pi\in\Pi(p,p')}
\int\d\x\,\d\bm y\,|\x-\bm y|^2\pi(\x,\bm y),
\eeq
where $\Pi(p,p')$ is the set of joint distributions $\pi(\x,\bm y)$ whose marginals are $p(\x)$ and $p'(\bm y)$.

Direct calculation of this distance is impractical in $N=74$ dimensions, and so we replace it with the Gelbrich lower bound\c{gelbrich1990formula}
\bea
\label{gelbrichmain}
G_s^2(t)&\equiv&|\bm m_s(t)|^2+Nm_2+{\rm Tr}\,C_s(t)\\
&&-2\sqrt{m_2}\,{\rm Tr}\!\left[C_s^{1/2}(t)\right]
\leq{\cal W}_2^2\!\left(p(\x,0),p_s(\x,t)\right), \nonumber
\eea
which depends only on the first two moments of $p$. In \eqq{gelbrichmain}, $m_2I$ is the covariance of the initial distribution $p(\x,0)$, where $I$ is the identity matrix and $m_2\equiv\int\d x\,x^2p_0(x)=0.234$. The mean of the initial distribution is zero. The mean $\bm m_s(t)$ and covariance $C_s(t)$ of the final distribution $p_s(\x,t)$ are defined in \eq{loadmoments}; here $C_s^{1/2}(t)$ is the positive-semidefinite square root of $C_s(t)$, and $|\bm m_s(t)|^2$ is the squared Euclidean norm of the mean.

Finally, we form the Gaussian lower bound $\eta_{\rm G}$ on the efficiency $\eta$:
\beq
\label{etadsmain}
\eta_{\rm G}(t)\equiv
\frac{\overline{G_s^2(t)}}
{\mu t \, \kt\,\overline{\Sigma_s^{\rm G}(t)}} \leq \eta(t) \leq 1.
\eeq

With our quantities of interest defined, we evaluate them numerically for different loading times $t_\ell$; additional simulation details and convergence checks are given in Appendix~\ref{app:simulation}. For each loading time $t_\ell$ we identify the observation time $t^*$ at which the test-set accuracy is greatest. Note that, for slow loading, readout can precede completion of loading: for example, $t^*=0.475$ for $t_\ell=0.5$. Thermodynamic quantities are evaluated up to the readout time.

In \f{fig2}(a) we show that slower loading reduces entropy production without impairing classification. We plot the mean entropy-production estimate $\overline{\Sigma^{\rm G}(t^*)}$ at the optimal observation time; entropy production decreases by more than an order of magnitude over the range of $t_\ell$ shown. The inset shows that classification accuracy $A(t^*)$ changes very little over this range, increasing slightly as loading rate decreases (the dashed line shows the corresponding result of the noise-free computer).

Gradual loading returns most of the available energy as work instead of dissipating it as heat. In panels (b) and (c) we show the work- and heat distributions over $10^4$ trajectories under three different loading rates. For a quench, the work distribution has zero mean, while finite-time loading shifts the work distribution to $W<0$, indicating that the computer does work on the loading source.

In \f{fig2}(d) we plot accuracy against the mean energy lost by the computer, $-\overline{\av{\Delta E_s(t)}}=\overline{\av{Q_s(t)-W_s(t)}}$, where the angle brackets denote an average over noise at fixed image and the overbar an average over data (note that the curves for $t_\ell\leq0.1$ almost coincide). Slower loading allows the computer to reach a fixed accuracy at smaller energy loss: accuracy is not fixed by the change of internal energy of the computer.

In \f{fig2}(e) we show the efficiency bound $\eta_{\rm G}(t)$ for each loading time. The readout times $t^\star$ corresponding to the observation time at which accuracy is greatest (shown inset in \f{fig2}(a)) are marked by colored dots. We observe an optimal loading time of $t_\ell=0.5$ at which the efficiency bound $\eta_{\rm G}$ is maximized; here, $\eta_{\rm G}=0.61$, indicating that the computer is operating within 40\% of the efficiency limit imposed by thermodynamics. For smaller loading rates ($t_\ell>0.5$) the longer runtime and smaller value of $\overline{G_s^2(t^\star)}$ outweigh the reduction in entropy production, while for larger loading rates ($t_\ell<0.5$) the shorter runtime fails to compensate for the increase in entropy production.

\subsection{Under the clock-acceleration procedure}

In this section we assess the efficiency of the computer under the clock-acceleration procedure of\cc{whitelam2025increasing}. Under this procedure the potential energy is uniformly rescaled and external noise is added in such a way that the stationary distribution of the computer is unchanged but its basic timescale is increased. We briefly recall the procedure before assessing the efficiency of the computer as it is applied. 

Within the system described by \eqq{lang1} we rescale the potential energy uniformly, sending $V(\x) \to \lambda V(\x)$, and add an independent Gaussian white noise of variance $2\mu\kt (\lambda-1)$. The resulting equation of motion is
\beq
\label{lang2}
\dot{\x}=\lambda\mu\ff(\x)+\sqrt{2\mu\kt}\,\xii(t)
+\sqrt{2\mu\kt(\lambda-1)}\,\zetaa(t),
\eeq
where the noises $\xii$ and $\zetaa$ are independent Gaussian white noises normalized as in \eq{lang1}. Their sum is also Gaussian, and so \eqq{lang2} is equivalent to
\beq
\label{lang3}
\dot{\x}=\lambda\mu\ff(\x)
+\sqrt{2 \lambda \mu \kB T}\,\xii(t).
\eeq

\eqq{lang3} can be regarded as a version of \eqq{lang1} with a renormalized mobility parameter, $\mu \to \lambda \mu \equiv \mu_\lambda$. Using \eqq{fp1}, the modified system can be seen to describe the same statistical ensemble of trajectories as the original, but on a renormalized timescale:
\beq
\label{corr}
p_{\lambda}(\x,t/\lambda)=p(\x,t).
\eeq
Here $p(\x,t)$ is the probability density of the original system (for which $\lambda=1$). Both potential rescaling and noise addition can be done experimentally\c{aifer2024thermodynamic,martinez2013effective,chupeau2018thermal,saha2023information}, and the clock-acceleration procedure was recently demonstrated in the laboratory using a colloidal particle in an optical trap\c{basak2026adding}. The procedure can be used to increase the basic rate of computation of a thermodynamic computer described by the overdamped Langevin equation, without changing the computer's program\c{whitelam2025increasing}.

We now examine how the speed limit \eq{sl} transforms under the clock-acceleration procedure. Specifically, we wish to compare the efficiency 
\beq
\label{eta2}
\eta(\tau) \equiv  \frac{{\cal W_2^2}(p(\x,0),p(\x,\tau))}{\mu \tau \kt \,\Sigma(\tau)} 
\eeq
of the original computer evaluated at time $\tau$ with the efficiency 
\beq
\label{eta3}
\eta_\lambda(\tau_\lambda) \equiv  \frac{{\cal W_2^2}(p_\lambda(\x,0),p_\lambda(\x,\tau_\lambda))}{\mu_\lambda \tau_\lambda \kt \,\Sigma_\lambda(\tau_\lambda)} 
\eeq
of the accelerated computer evaluated at time $\tau_\lambda \equiv \tau/\lambda$ (here, for clarity, we suppress data labels).

The numerators of \eq{eta2} and \eq{eta3} are identical. From Equation~\eq{corr} it follows that the computational progress of the original computer at time $\tau$ is equal to that of the accelerated computer at time $\tau/\lambda$, and so
\beq
\label{wscale}
{\cal W}_2^2[p_\lambda(\x,0),p_\lambda(\x,\tau_\lambda)]
={\cal W}_2^2[p(\x,0),p(\x,\tau)].
\eeq

The entropy production of the two computers is also identical: under acceleration, the instantaneous rate of entropy production increases by a factor of $\lambda$, but the run time diminishes by the same factor. To see this, note that the velocity current transforms under clock acceleration as $\nuu_\lambda(\x,t/\lambda)=\lambda\nuu(\x,t)$, and so the instantaneous rate of entropy production is
\bea
\label{iep}
\sigma_\lambda(t/\lambda)
&=&\frac{1}{\mu_\lambda\kt}
\int\d\x\,\|\nuu_\lambda(\x,t/\lambda)\|^2
p_\lambda(\x,t/\lambda)\nonumber\\
&=&\frac{\lambda^2}{\mu_\lambda\kt}
\int\d\x\,\|\nuu(\x,t)\|^2
p(\x,t)\nonumber\\
&=&\lambda\sigma(t).
\eea
The total entropy produced in time $\tau_\lambda$ by the accelerated computer is then
\beq
\label{sigmascale}
\Sigma_\lambda
=\int_0^{\tau_\lambda}\!\d u\,\sigma_\lambda(u)=\int_0^{\tau/\lambda}\!\d u\,\lambda\sigma(\lambda u)
=\int_0^\tau\!\d t\,\sigma(t)=\Sigma,
\eeq
which is the same as that of the original computer. 

The remaining factors to consider in the denominator of \eq{eta2} and \eq{eta3} are the product of mobility and runtime, and these are also the same: $\mu_\lambda \tau_\lambda = \mu \tau$.

Thus the efficiencies \eq{eta2} and \eq{eta3} are the same. The clock-acceleration procedure does not affect the efficiency of a thermodynamic computer: there is no thermodynamic proscription against a faster runtime at fixed entropy production.

However, clock acceleration {\em does} have a thermodynamic cost: it increases the heat emission and the instantaneous power consumption of the computer. This is most clearly seen if we consider \eqq{lang3} to describe a system with the original mobility $\mu$ and scaled potential $\lambda V$, in contact with effective thermal bath of temperature $\lambda T$ (an alternative thermodynamic picture, in which the injected noise is considered to be an external force, is discussed in the SI of\cc{whitelam2025increasing}). From the definition of heat \eq{heat} we can show that the total heat emitted and the mean dissipated power transform under the clock-acceleration procedure as
\beq
\label{clockheat}
\av{Q_\lambda(\tau_\lambda)}=\lambda\av{Q(\tau)},\quad
\frac{\av{Q_\lambda(\tau_\lambda)}}{\tau_\lambda}
=\lambda^2\frac{\av{Q(\tau)}}{\tau},
\eeq
which are larger $(\lambda > 1)$ than for the original computer (for which $\lambda=1$). Under the clock-acceleration procedure, the usual statement that faster operation requires greater dissipation refers to heat, not to total entropy production.

In \f{fig2} we confirm these statements numerically.  \f{fig2}(f) shows the accuracy as a function of time for the thermodynamic computer described by \eqq{pot} and \eqq{lang2}, for $\lambda=1,2,4,$ and $8$. Increasing $\lambda$ compresses the dynamics in time, but does not change the trajectory distribution. As a result, the computer's inference accuracy is unchanged. The inset to the figure confirms that the heat emitted up to time $\tau_\lambda$ increases in proportion to $\lambda$, while the total entropy production is unchanged. 

Panel (g) shows the data-averaged Gaussian efficiency bound on $\eta$,
\beq
\label{eta4}
\eta_{{\rm G},\lambda}(\tau_\lambda)
\equiv\frac{\overline{G_{\lambda,s}^2(\tau_\lambda)}}
{\mu_\lambda\tau_\lambda\kt\,
\overline{\Sigma_{\lambda,s}^{\rm G}(\tau_\lambda)}},
\eeq
which confirms numerically the analytic results of this section: the efficiency curves of the computers run at different values of $\lambda$ are time-rescaled versions of each other. As a result, clock acceleration permits inference at the same test-set accuracy and thermodynamic efficiency, but with shorter run time. The thermodynamic cost required to achieve this speed up is increased mean power and heat emission.

\section{Conclusions}
\label{sec:conclusions}

We have described the operation and thermodynamics of a model thermodynamic computer. The computer classifies MNIST images using a single nonequilibrium dynamical trajectory per image, and does so with an accuracy comparable to that of a simple multilayer perceptron. The thermodynamic efficiency of the computer can be varied through choice of inference protocol, without significant change in operational accuracy.

Loading the computer's interactions gradually reduces heat and entropy production, and allows an increase of the efficiency of the computer to within 40\% of the thermodynamic limit. The cost of this increase is greater run time.

Under the clock-acceleration procedure of\cc{whitelam2025increasing} the computer can be run faster at the same efficiency, with no increase in entropy production. The cost of this procedure is increased dissipated heat and mean power.

These results complement the energy-time-accuracy tradeoffs identified in\cc{rolandi2026energy}, and the finding that thermodynamic logic gates can operate in a range of dissipative regimes\c{whitelam2026evolutionary}. A thermodynamic computer trained for a particular task can perform it under a wide range of thermodynamic conditions.

\section{Acknowledgements} This work was done at the Molecular Foundry, supported by the Office of Science, Office of Basic Energy Sciences, of the U.S. Department of Energy under Contract No. DE-AC02-05CH11231. In writing this paper I used AI tools (Anthropic's Claude Opus 5 and Fable 5, and OpenAI's GPT-5.6 Sol) to produce batched Python code using the handwritten serial C++ code of\cc{whitelam2026training} as a starting point. I used the same AI tools to produce simulation scripts and to compile the figures.

\appendix
\section{Simulation details}
\label{app:simulation}
\renewcommand{\thefigure}{A\arabic{figure}}
\renewcommand{\theHfigure}{A\arabic{figure}}
\setcounter{figure}{0}

We integrated the Langevin equations using the Euler--Maruyama method\c{kloeden1992numerical}. Writing $t_n=n\Delta t$, one step of the dynamics \eq{langload} corresponds to the update
\beq
\label{em}
\x_{n+1}=\x_n-\mu\nabla V_s(\x_n,t_{n+1})\Delta t
+\sqrt{2\mu\kt\Delta t}\,\bm z_n,
\eeq
where the components of $\bm z_n$ are independent standard normal variables. Times are measured in units of $\mu^{-1}$.

We verified that the results of the paper are insensitive to various numerical and sampling choices:
\begin{itemize}
\item The data in \f{fig1}(b--h) were obtained using an integration timestep $\Delta t=5\times10^{-3}$, except for the noise-free curve in panel (g), which uses $\Delta t=10^{-3}$. We verified that the test-set accuracies calculated using the two timesteps are essentially identical. Panel (h) was calculated using the first 5000 test-set images.
\item The data in \f{fig2}(a--e) were obtained using an integration timestep $\Delta t=10^{-3}$. 
\item The data in \f{fig2}(f,g) were obtained using an integration timestep $\Delta t=10^{-3}/\lambda$.
\item We estimated the components of $\eta_{\rm G}$ using $R=2000$ trajectories for each of 50 test images, five per digit class, weighted by the digit frequencies of the full test set. The mean energy loss in \f{fig2}(d) is estimated using the same sample; the accuracies and the work and heat distributions in panels (b,c) use all $10^4$ test images. We removed the leading finite-$R$ bias of $G_s^2$ by linear extrapolation in $1/R$. At the most accurate readout times for the loading times $t_\ell=0,0.1,$ and $0.5$, we find that quadrupling the number of images, doubling $R$, or halving $\Delta t$ changes $\eta_{\rm G}$ by at most $0.003$, and changes $\overline{\Sigma_s^{\rm G}}$ and $\overline{G_s^2}$ by less than $0.6\%$.
\end{itemize}

 Each MNIST pixel value was divided by 255, then shifted and scaled so that training-set data have zero pixel mean and unit variance.  Each image was then rescaled to have the same Euclidean pixel norm. 

The efficiency $\eta_{\rm G}$ in \eqq{etadsmain} is the ratio of the data-averaged transport and thermodynamic cost. We also calculated a modified efficiency, defined as the average over data of the per-image ratio: 
\beq
\label{etaimage}
\tilde{\eta}_{\rm G}(t)
\equiv
\frac{1}{\mu t\kt} \overline{\left(\frac{G_s^2(t)}{\Sigma_s^{\rm G}(t)}\right)}.
\eeq
We found the modified efficiency (evaluated under loading and clock acceleration) to be the same as the original efficiency to within a value $1.1\times10^{-3}$.


\def\doibase#110.{https://doi.org/10.}

\end{document}